\documentclass[journal=jpcafh,manuscript=article]{achemso} 
\newcommand{\rvec}{\mathrm {\mathbf {r}}} 
 
\newcommand{\Rvec}{\mathrm {\mathbf {R}}} 
\newcommand{\Cvec}{\mathrm {\mathbf {C}}} 
\newcommand{\Svec}{\mathrm {\mathbf {S}}} 
\newcommand{\Ivec}{\mathrm {\mathbf {I}}} 
\newcommand{\kvec}{\mathrm {\mathbf {k}}}

\usepackage{graphicx}
\usepackage{subfigure}
\usepackage{xcolor}
\usepackage{enumerate}
\usepackage{braket}
\usepackage{amsmath, bm}
\usepackage{stackengine}

\usepackage{color, soul}
\definecolor{darkblue}{rgb}{0,0,0.5}
\setulcolor{darkblue}

\author{Abhisek Ghosal, Tanmay Mandal}
\author{Amlan K.~Roy}
\affiliation{Department of Chemical Sciences\\
Indian Institute of Science Education and Research (IISER) Kolkata \\  
Nadia, Mohanpur-741246, WB, India}
\email{akroy@iiserkol.ac.in, akroy6k@gmail.com.}                                           
\title{Efficient HF exchange evaluation through Fourier convolution in Cartesian grid for orbital-dependent density
functionals}

\begin{document}
\begin{abstract}
We present a purely numerical approach in Cartesian grid, for efficient computation of Hartree-Fock (HF) exchange contribution
in the HF and density functional theory models. This takes inspiration from a recently developed algorithm [Liu \emph{et al.}, 
J.~Chem.~Theor.~Comput.~ \textbf{13}, 2571 (2017)]. A key component is the accurate evaluation of electrostatic potential integral, 
which is the rate-determining 
step. This introduces the Fourier convolution theorem in conjunction with a range-separated Coulomb interaction kernel. The 
latter is efficiently mapped into real grid through a simple optimization procedure, giving rise to a constraint in the 
range-separated parameter. The overall process offers logarithmic scaling with respect to molecular size. It is then 
extended towards global hybrid functionals such as B3LYP, PBE0 and BHLYP within pseudopotential Kohn-Sham theory, through 
an LCAO-MO ansatz in Cartesian grid, developed earlier in our laboratory. For sake of comparison, a parallel semi-numerical 
approach has also been worked out that exploits the familiar Obara-Saika recursion algorithm. An excellent agreement between 
these two routes is demonstrated through total energy and orbital energy in a series of atoms and molecules (including 10 
$\pi$-electron molecules), employing an LANL2DZ-type basis function. A critical analysis of these two algorithms reveals that the 
proposed numerical scheme could lead to very attractive and competitive scaling. The success of our approach also enables us for 
further development of optimally tuned range-separated hybrid and hyper functionals.  
\end{abstract}

\section{1. Introduction}
Kohn-Sham density functional theory (KS-DFT) \citep{hohenberg64,kohn65} has been found to be an indispensable tool for 
determining structure and properties of many-electron systems like atoms, molecules, clusters, nano-materials and periodic 
systems, for past several decades\citep{cohen11, burke12, becke14, jones15}. The KS mapping is exact, \emph{in principle}, 
and has the ability to capture many-body correlation effects at a computational cost comparable to Hartree-Fock (HF) theory. 
But a major bottleneck lies in the exchange-correlation (XC) functional that uses electron density, $\rho(\rvec)$, to 
describe non-classical effects under the mean-field formalism of KS scheme. As its exact form is unknown as yet, it has 
to be approximated and accordingly, $ E_{\mathrm{xc}}^{\mathrm{approx}}[\rho(\rvec)] = 
\int d^3\rvec f^{\mathrm{approx}}_{\mathrm{xc}}(\rho,\nabla\rho,\tau,e^{\mathrm{x}},\dots)$. Commonly used functionals 
can be hierarchically characterized using these variables. Thus starting from a simpler local density approximation (LDA) 
(containing $\rho$ only), to general gradient approximation (GGA) (with addition of gradient of electron density, 
$\nabla\rho$), to meta-GGA (further addition of kinetic energy density, $\tau$), one gets a progressively involved
and sophisticated range of functionals. With further inclusion of exact (also called HF) exchange energy, one gets the hybrid 
functionals, while incorporation of exact exchange energy density, $e^{\mathrm{x}}$, leads to a hyper-GGA type XC 
functionals, residing in the fourth rung of Jacob's ladder. Now we also have functionals which go beyond the 
fourth rung (and include virtual orbitals), thus requiring higher computational cost, and have been successfully 
implemented in literature in the context of various physio-chemical properties\citep{grimme07, zhang09}.

In general the approximate local or semi-local functionals experience certain issues regarding (i) correct asymptotic 
behavior of exact exchange potential at long range and (ii) non-cancellation of spurious Coulomb self-repulsion 
energy--the so-called self-interaction error (SIE)\citep{perdew81, bao18}. These two points are connected to each other and may be 
dealt with by bringing exact exchange in to the picture. This component may be combined with semi-local functionals, 
empirically\citep{becke93b} or non-empirically \citep{perdew96a, guido13}, thus improving the asymptotic nature, and 
consequently reducing the SIE to a greater extent. Accordingly, the global hybrid functionals \citep{becke93b} (B3LYP being a
prototypical model) were proposed in the literature\citep{adamo99, ernzerhof99, zhao08}, which improved the chemical accuracy 
in a large number of physico-chemical properties\citep{junior08, mangiatordi12}. On the other hand, non-dynamic (or static) correlation 
seems to be a serious problem for current DFT functionals\citep{becke14} which arises when an electronic state cannot be properly 
described by a single Slater determinant. It requires a linear combination of degenerate or nearly degenerate Slater 
determinants. Naturally it is very much demanding to model this correlation accurately within a single determinantal
KS-DFT framework, and some methods have been proposed in the past decade under the name of hyper-GGA functionals\citep{becke03, 
becke05, perdew08}. More recently, general single-determinant model functionals have also been put forth to treat this  
correlation up to the dissociation limit of a co-valent bond\citep{becke13a, kong15}. All these emerging functionals 
belonging to hyper-GGA category, require $e^{\mathrm{x}}$ as a fundamental variable and in general, their computational cost 
is much higher compared to the global hybrid functionals. 

The HF exchange energy and corresponding matrix elements are usually evaluated through four-center electron repulsion 
integrals (ERI) analytically, within a Gaussian basis expansion of molecular orbitals (MO), providing following 
contributions to KS-Fock matrix, 
\begin{equation} 
F_{\mu\nu}^{\mathrm{x}\sigma}=\sum_{\lambda\eta} P_{\lambda\eta}^{\sigma} (\mu\lambda|\eta\nu) = \sum_{\lambda\eta} 
P_{\lambda\eta}^{\sigma} \int \int \frac{\chi_{\mu,\sigma}(\rvec) \chi_{\lambda,\sigma}(\rvec)\chi_{\nu,\sigma}
(\rvec^{\prime})\chi_{\eta,\sigma}(\rvec^{\prime})}{|\rvec-\rvec^{\prime}|} d\rvec d\rvec^{\prime}. 
\end{equation}
Here $(\mu\lambda|\eta\nu)$ is the four center ERI with $\mu, \nu, \lambda, \eta$ representing the atomic orbitals (AO) or 
contracted basis, and $P_{\lambda\eta}^{\sigma}$ is an element of the single-particle spin density matrix, $P^{\sigma}$ 
having spin $\sigma$. Throughout the past few decades, there has been extensive methodological developments to optimize 
its evaluation using a range of approaches varying in complexity, sophistication and accuracy. It is worthwhile 
mentioning some of the notable ones, such as pseudo-spectral scheme using Gaussian basis functions, multipole 
accelerated algorithm, LinK, truncated or short-range exchange kernels with/without the use of resolution-of-the-identity 
(RI) approximation, chain-of-sphere exchange based on quadrature, auxiliary density matrix, plane-wave basis functions, 
metropolis stochastic, pair-atomic RI, adaptively compressed exchange operator, density fitting with local domains, 
real-space with projection operators, rigorous integral screening, occupied-orbital fast multipole \citep{chasman98, 
ochsenfeld00, neese09, cytter14, manzer15, lin16, boffi16, thompson17, le18}, etc. The HF exchange energy density can be  
represented as, 
\begin{eqnarray}
e^{\mathrm{x}}_{\sigma}(\rvec) = -\sum_{i}^{occ} \sum_{j}^{occ} \int \frac{\phi^{\star}_{i,\sigma}(\rvec)
\phi^{\star}_{j,\sigma}(\rvec)\phi_{i,\sigma}(\rvec^{\prime})\phi_{j,\sigma}(\rvec^{\prime})}{|\rvec-\rvec^{\prime}|}
d\rvec^{\prime} \nonumber \\
=-\sum_{\mu\nu}\sum_{\lambda \eta} P_{\mu\nu}^{\sigma} P_{\lambda\eta}^{\sigma}\int \frac{\chi_{\mu,\sigma}(\rvec) 
\chi_{\lambda,\sigma}(\rvec)\chi_{\nu,\sigma}(\rvec^{\prime})\chi_{\eta,\sigma}(\rvec^{\prime})}{|\rvec-\rvec^{\prime}|}
d\rvec^{\prime}. 
\end{eqnarray}
The two expressions are defined in terms of KS occupied MO ($\phi_{\sigma}$) and AO ($\chi_{\sigma}$) respectively. 
Complex conjugate sign is omitted here since density matrix and basis are generally in real form. At  
first glance, it seems to be computationally more expensive than normal exchange energy calculation due to the fact that 
it needs to be evaluated \emph{at each grid point} with four AO indices. But some recent developments proposed in the 
literature show substantial computational cost reduction via pair-atomic RI approximation\citep{proynov10, proynov12, liu12} 
or using a  semi-numerical (SNR) scheme\citep{kossmann09, bahmann15, laqua18, liu17}. Here we mainly focus on the latter 
implementation, where one of the integrations involved in ERI is carried out numerically on the effective spatial grid. 

The main motivation of this work is thus to calculate HF exchange density, energy and its matrix elements which are essential 
ingredients for development of some of the new-generation (in particular, orbital-dependent) XC functionals. This is 
performed accurately in real-space Cartesian coordinate grid (CCG) following the procedure \citep{liu17}, where a key step is the
identification and evaluation of an intermediate quantity, namely, \emph{two-center electrostatic potential} (ESP), defined as, 
\begin{equation}
v_{\nu\eta}(\rvec)=\int\frac{\chi_{\nu}(\rvec^{\prime})\chi_{\eta}(\rvec^{\prime})}
{|\rvec-\rvec^{\prime}|}d\rvec^{\prime} 
=\sum_{p}\sum_{q}\int\frac{\varphi_{\nu}^{p}(\rvec^{\prime})\varphi_{\eta}^{q}(\rvec^{\prime})}
{|\rvec-\rvec^{\prime}|}d\rvec^{\prime}. 
\end{equation}
The first expression is defined in terms of AO, whereas the second one in terms of primitive basis functions, $\varphi_{\mu}^p$ 
for a given $\chi_{\mu}$ (for simplicity we remove the spin indices). Note that symbols $\mu, \nu$ signify contracted basis 
functions, whereas $p,q$ label primitive functions. We propose a direct, efficient \emph{numerical} (NR) scheme based 
on \emph{Fourier convolution theorem} (FCT) for accurate estimation of this ESP integral (first expression) using a 
range-separation (RS) technique, corresponding to long-and short-range interactions for \emph{Coulomb interaction 
kernel}. A major
concern is how one can define the optimal RS parameter for a successful mapping of this interaction kernel in CCG 
from \emph{first principles} rather than empirically. Towards this direction, we have employed a recently developed grid 
optimization procedure \citep{ghosal16, ghosal18} with respect to total energy in non-uniform CCG (originally used for  
evaluation of classical Hartree potential) to optimize the RS parameter, to a very good level of accuracy, using a well-defined 
constraint, suggested in the literature. The successful implementation of this approach also paves the way for further development 
of the so-called range-separated hybrid functionals in connection with generalized KS theorem\citep{seidl96}. The 
underlying algorithm of the SNR method and subsequently a detailed account of our proposed NR scheme, denoted as 
RS-FCT, for estimation of this ESP integral, is provided in Sec.~2. Then its implementation is carried out using 
a pseudopotential KS-DFT in CCG, developed in our lab\citep{roy08, roy08a, roy11, ghosal16, ghosal18}. The relevant cost analysis 
between these two routes is also sketched upon. Section 3 offers the necessary computational and technical details. Finally, the 
feasibility, performance and accuracy of 
our results are critically assessed through quantities such as total energy and orbital energy, in Sec.~4, employing 
three-types of global hybrid functionals B3LYP"\citep{becke93a}, PBE0\citep{perdew96a} and BHLYP\citep{becke93b} along with 
the HF calculation for a decent (total 40) number of atoms and molecules. Some concluding remarks as well as the future prospect
is given in Sec.~4.

\section{2. Methodology}
\subsection{2.1. HF exchange energy, density and matrix elements}
Let us begin with the SNR approach, where HF exchange energy is computed numerically by integrating the corresponding density, 
$e^{\mathrm{x}}_{\sigma} (\rvec)$ as given by, 
\begin{equation}
E_{\sigma}^{\mathrm{x}}=\frac{1}{2}\int e_{\sigma}^{\mathrm{x}}(\rvec) d\rvec.
\end{equation}
Now recasting Eq.~(2) in the following form,  
\begin{equation}
e_{\sigma}^{\mathrm{x}}(\rvec)=-\sum_{\nu}Q_{\nu}^{\sigma}(\rvec)M_{\nu}^{\sigma}(\rvec), 
\end{equation}
one may envisage the construction of $e^{\mathrm{x}}_{\sigma} (\rvec)$ in three steps of comparable computational cost. The first 
quantity, $Q_{\nu}^{\sigma}(\rvec)$ may be represented as follows,  
\begin{equation}
Q_{\nu}^{\sigma}(\rvec) = \sum_{\mu} \chi_{\mu,\sigma}(\rvec) P_{\mu\nu}^{\sigma},
\end{equation}  
in which the density matrix is combined with AOs through a simple matrix multiplication. The computational cost of this step 
scales as $\mathcal{O}(N_gN_{B}^{2})$, with $N_g, N_B$ denoting total number of grid points, and number of AO basis. Next 
crucial (rate-determining) step is to evaluate the ESP integral (embedded in $M_{\nu}^{\sigma}(\rvec)$), which according to Eq.~(3) 
(first expression), also scales as  $\mathcal{O}(N_gN_{B}^{2})$. Alternately one may perform this integral analytically 
(corresponding to second expression of Eq.~(3)) using primitive functions and standard recursion algorithm \citep{obara86, liu16}. 
In that scenario, evidently \emph{each ESP integral} scales as $\mathcal{O}(N_gN_{P}^{2})$, with $N_{P}$ 
identifying average number of primitive functions. The final step consists of computation of the quantity, 
$M_{\nu}^{\sigma}(\rvec)$ in accordance with the following expression, 
\begin{equation}
M_{\nu}^{\sigma}(\rvec)=\sum_{\eta}Q_{\eta}^{\sigma}(\rvec)v_{\nu\eta}^{\sigma}(\rvec).
\end{equation}
The scaling of this step is also same as that of ESP integral evaluation, but requires fewer steps than the latter; needing
only one multiplication and one addition at innermost loop. This effectively provides an NR way to
compute the HF exchange matrix, which according to Eq.~(1), can be rewritten as,
\begin{equation}
\frac{\partial{E^{x}_{\sigma}}}{\partial{P_{\lambda\eta}^{\sigma}}}=F_{\mu\nu}^{\mathrm{x}\sigma}=-\int_{\rvec}\chi_{\mu}
(\rvec)M_{\nu}^{\sigma}(\rvec)d\rvec.
\end{equation} 

\subsection{2.2. Pseudopotential KS-DFT in CCG}
For a many-electron system, one can write the single-particle KS equation in presence of pseudopotential as (atomic 
unit employed unless stated otherwise), 
\begin{equation}
\bigg[ -\frac{1}{2} \nabla^2 + v^{\mathrm{p}}_{\mathrm{ion}}(\rvec) + v_{\mathrm{ext}}(\rvec) + v_{\mathrm{h}}[\rho(\rvec)] 
+ v_{\mathrm{xc}}[\rho(\rvec)] \bigg] \psi_i^{\sigma}(\rvec) = \epsilon_i \psi_i^{\sigma}(\rvec),
\end{equation}
where $v^{\mathrm{p}}_{\mathrm{ion}}$ denotes the ionic pseudopotential, written as below, 
\begin{equation}
v^{\mathrm{p}}_{\mathrm{ion}}(\rvec) = \sum_{\Rvec_a} v^{\mathrm{p}}_{\mathrm{ion},\mathrm{a}} (\rvec-\Rvec_a).
\end{equation}
In the above equation, $v^{\mathrm{p}}_{\mathrm{ion},\mathrm{a}}$ signifies ion-core pseudopotential associated with atom A, 
situated at $\Rvec_a$. The classical Coulomb (Hartree) term, $v_{\mathrm{h}}[\rho(\rvec)]$ describes usual electrostatic 
interaction amongst valence electrons whereas $v_{\mathrm{xc}}[\rho(\rvec)]$ signifies the non-classical XC part of latter, 
and $\{ \psi^{\sigma}_{i},\sigma= \alpha \quad \mathrm{or} \quad  \beta \}$ corresponds to a set of $N$ occupied orthonormal 
spin molecular orbitals (spin-MOs). Within LCAO-MO approximation, the coefficients for expansion of spin-MOs satisfy a 
set of equations, very similar to that in HF theory,  
\begin{equation}
\sum_{\nu} F_{\mu \nu}^{\sigma} C_{\nu i}^{\sigma} = \epsilon_{i}^{\sigma} \sum_{\nu}^{\sigma} S_{\mu \nu} 
C_{\nu i}^{\sigma},
\end{equation}
satisfying the orthonormality condition, $(\Cvec^{\sigma})^{\dagger} \Svec \Cvec^{\sigma} = \Ivec$. Here $\Cvec^{\sigma}$ 
contains the respective spin-MO coefficients $\{C_{\nu i}^{\sigma}\}$ for a given spin-MO $\psi_i^{\sigma}(\rvec)$, $\Svec$ is 
the usual overlap matrix corresponding to elements $S_{\mu \nu}$, $\bm{\epsilon}^{\sigma}$ refers to diagonal matrix of 
spin-MO eigenvalues $\{\epsilon_{i}^{\sigma}\}$. The KS-Fock matrix has elements $F_{\mu \nu}^{\sigma}$, constituting of 
following contributions, 
\begin{equation}
F_{\mu \nu}^{\sigma} = H_{\mu \nu}^{\mathrm{core}} + J_{\mu \nu} + F_{\mu \nu}^{\mathrm{xc}\sigma}.
\end{equation} 
In this equation, all one-electron contributions, such as kinetic energy, nuclear-electron attraction and pseudopotential matrix 
elements are included in first term, whereas $J_{\mu \nu}$ and $F_{\mu \nu}^{\mathrm{xc}\sigma}$ account for classical 
Hartree and XC potentials respectively. 
 
Now we discretize various quantities like localized basis function, electron density, MO as well as two-electron potentials 
directly on a 3D cubic box having $x,y,z$ axes, 
\begin{eqnarray}
r_{i}=r_{0}+(i-1)h_{r}, \quad i=1,2,3,....,N_{r}~, \quad r_{0}=-\frac{N_{r}h_{r}}{2}, \quad  r \in \{ x,y,z \},
\end{eqnarray}
where $h_{r}, N_r$ denote grid spacing and total number of points along each directions respectively. The electron density 
$\rho(\rvec)$ in this grid may be simply written as (``g" symbolizes discretized grid),
\begin{equation}
\rho(\rvec_g) = \sum_{\mu,\nu} P_{\mu \nu } \chi_{\mu}(\rvec_g) \chi_{\nu}(\rvec_g). 
\end{equation}
At this stage, the two-electron contributions of KS matrix are computed through direct numerical integration in the grid, 
\begin{equation}
F_{\mu \nu}^{\mathrm{hxc}} = \langle \chi_{\mu}(\rvec_g)|v_{\mathrm{hxc}}(\rvec_g)|\chi_{\nu}(\rvec_g) \rangle = h_x h_y h_z 
\sum_g \chi_{\mu}(\rvec_g) v_{\mathrm{hxc}}(\rvec_g) \chi_{\nu}(\rvec_g).
 \end{equation}
where $v_{hxc}$ refers to the combined Hartree and XC potential. The detailed construction of various potentials in CCG has 
been well documented in our earlier work\citep{roy08, roy08a, roy11, ghosal16, ghosal18}; hence not repeated here.
Similarly, the HF exchange energy and its contributions towards KS-Fock matrix can be computed 
numerically in CCG as, 
\begin{eqnarray}
E_{\sigma}^{\mathrm{x}}=  \frac{1}{2} h_x h_y h_z \sum_g e_{\sigma}^{\mathrm{x}}(\rvec_g), \nonumber \\
F_{\mu\nu}^{\mathrm{x}\sigma}=- h_x h_y h_z \sum_g \chi_{\mu}(\rvec_g)M_{\nu}^{\sigma}(\rvec_g).
\end{eqnarray}

\subsection{2.3. RS-FCT for computation of ESP integral in CCG}
Let us start with two functions $f(\rvec)$ and $F(\kvec)$ in $\rvec$ and $\kvec$ spaces, which are Fourier transforms (FT) 
(denoted by $\mathcal{F}$) of each other, as given below,  
\begin{eqnarray}
f(\rvec) & = & \mathcal{F}(\kvec) =  \int_{-\infty}^{\infty} F(\kvec) e^{2\pi i \rvec\kvec} d\kvec \nonumber \\
F(\kvec) & = & \mathcal{F}^{-1} (\rvec)=  \int_{-\infty}^{\infty} f(\rvec) e^{-2\pi i \rvec\kvec} d\rvec. 
\end{eqnarray}
With two functions $f(\rvec), g(\rvec)$, one can form the 
convolution, defined by,  
\begin{equation}
f\star g = \int_{-\infty}^{\infty} f(\rvec^{\prime}) g(\rvec-\rvec^{\prime}) d\rvec^{\prime}.
\end{equation}
The convolution theorem states that FT of convolution is just the product of individual FTs, 
\begin{eqnarray}
\mathcal{F}(f \star g) & = & F(\kvec).G(\kvec) \nonumber \\
f \star g & = & \mathcal{F}^{-1}(F(\kvec).G(\kvec)).
\end{eqnarray}
One can make use of this theorem to rewrite the ESP integral as, 
\begin{equation}
v_{\nu\eta}(\rvec)=\int\frac{\chi_{\nu}(\rvec^{\prime})\chi_{\eta}(\rvec^{\prime})}{|\rvec-\rvec^{\prime}|}d\rvec^{\prime}
=\int\frac{\chi_{\mu\nu}(\rvec^{\prime})}{|\rvec-\rvec^{\prime}|}d\rvec^{\prime}=\chi_{\nu\eta}(\rvec)\star v^{c}(\rvec).
\end{equation}
The spin indices have been removed for simplicity. The last expression is in terms of convolution integral, 
where $\chi_{\nu\eta}$ denotes simple multiplication of two AO basis and $v^{c}(\rvec)$ represents Coulomb interaction 
kernel. Now one can invoke FCT to further simplify this integral,  
\begin{equation}
v_{\nu\eta} (\rvec) =\mathcal{F}^{-1}(v^{c}\{\kvec) \chi_{\nu\eta}(\kvec)\} \quad \mathrm{where} \quad \chi_{\nu\eta}(\kvec)
=\mathcal{F}\{\chi_{\nu\eta}(\rvec)\}. 
\end{equation}
Here $v^{c}(\kvec)$ and $\chi(\kvec)$ stand for Fourier integrals of Coulomb kernel and AO basis respectively. The main 
concern lies in an 
accurate mapping of the former, which has singularity at $\rvec \to 0$. In order to alleviate this problem, we apply a 
simple RS technique from the works of \citep{gill96, gill96a}, expanding the kernel into long- and short-range components 
with a suitably chosen RS parameter ($\zeta$). This gives rise to the following expression,  
\begin{eqnarray}
v^{c}(\rvec)=\frac{\mathrm{erf}(\zeta\rvec)}{\rvec}+\frac{\mathrm{erfc}(\zeta\rvec)}{\rvec} \nonumber \\
v^{c}(\rvec_g)= v^{c}_{long}(\rvec_g) + v^{c}_{short}(\rvec_g). 
\end{eqnarray}
In the above equation, $\mathrm{erf}(x)$ and $\mathrm{erfc}(x)$ denote error function and its complement respectively, while 
the second expression is written in real-space CCG. The Fourier integral of Coulomb kernel can be separated out as follows: 
(i) FT of short-range part is treated analytically and (ii) long-range portion is computed directly from FFT of 
corresponding real-space CCG values. Then the remaining problem lies in finding an optimum value of parameter 
$\zeta_{\mathrm{opt}}$ for successful mapping of Coulomb kernel in CCG from \emph{first principles}. In this regard we 
propose a simple procedure which is found to be sufficiently accurate for all practical
purposes (as exemplified in results that follows). This prompts us to write, 
\begin{equation}
\zeta_{\mathrm{opt}} \equiv \underset{\zeta }{\mathrm{opt}} \ E_{\sigma}^{\mathrm{x}} \equiv \underset{\zeta }{\mathrm{opt}} \ 
E_{\mathrm{tot}} = \underset{N_x,N_y,N_z} {\mathrm{opt}} E_{\emph{tot}}, \quad \mathrm{at \ fixed} \ h_{\rvec},
\end{equation}
using a suitably defined constraint \cite{martyna99} $(\zeta \times L =7 )$, where $L\ (= N_\rvec h_\rvec; \ \rvec \in 
\{x,y,z\})$ 
is the smallest side of simulating box. This technique was already implemented in grid optimization procedure 
with respect to total energy in a non-uniform CCG \citep{ghosal16,ghosal18}. 

From the foregoing discussion it is clear that, each ESP integral involves only a combination of FFT (two forward 
and one backward transformation simultaneously) and hence scales as $\mathcal{O}(N_{g} \log N_g)$. It is to be noted 
that this prescription gives a logarithmic scaling with grid size, in contrast to the quadratic scaling (with respect 
to primitive basis functions, $N_P$) in SNR scheme. 

\section{3. Computational details}
The above described strategy for HF exchange component is implemented in case of three global hybrid functionals, namely, 
B3LYP, PBE0 and BHLYP, containing a variable amount of former with conventional DFT XC functional. As such, the XC 
energy corresponding to these three functionals are defined as follows, 
\begin{equation}
E_{\mathrm{xc}}^{\mathrm{B3LYP}} = (1-a_0)E^{\mathrm{x}}_{\mathrm{LSDA}}+a_0 E^{\mathrm{x}}_{\mathrm{HF}}+ 
a_{\mathrm{x}} E^{\mathrm{x}}_{\mathrm{B88}}+a_{\mathrm{c}} E^{\mathrm{c}}_{\mathrm{LYP}} + (1-a_{\mathrm{c}})
E^{\mathrm{c}}_{\mathrm{VWN}},
\end{equation}
\begin{equation}
E_{\mathrm{xc}}^{\mathrm{PBE0}} = b_0 E^{\mathrm{x}}_{\mathrm{HF}} + (1-b_0)E^{\mathrm{x}}_{\mathrm{PBE}} + 
E^{\mathrm{c}}_{\mathrm{PBE}}, 
\end{equation}
and
\begin{equation}
E_{\mathrm{xc}}^{\mathrm{BHLYP}} = (1-c_0)E^{\mathrm{x}}_{\mathrm{LSDA}}+ 
c_0 E^{\mathrm{x}}_{\mathrm{HF}}+ E^{\mathrm{c}}_{\mathrm{LYP}}.
\end{equation}
Following the work of Stephens \emph{et al.}, \citep{stephens94} values of $a_0, a_{\mathrm{x}}, a_{\mathrm{c}}$ 
are $0.2$, $0.72$, $0.81$ for B3LYP, whereas in case of PBE0\citep{perdew96a}, $b_0=0.25$. Note that the amount of HF
exchange contribution in PBE0 is slighter higher than that of B3LYP, but even larger amount ($c_0=0.5$) is attributed in 
BHLYP. These are computed using KS orbitals obtained from solution of Eq.~(9) in real-space CCG through the procedure 
delineated in previous subsections.  

\begingroup                      
\begin{table}[t]      
\caption{\label{tab:table1} Convergence of HF energy$^{{\dagger},\ddagger}$ of HCl ($R=1.275 \textup{\AA}$) in the grid 
($h_r=0.3$). All results are in a.u.}
\begin{tabular} {ccclcccrl}
\cline{1-9}
 \multicolumn{4}{c}{Set~I }  & & \multicolumn{4}{c}{Set~II}      \\
\cline{1-4} \cline{6-9} 
$N_x$   &   $N_y$   &    $N_z$  &  $ \langle E \rangle $ &    & $N_x$   &   $N_y$   &    $N_z$  &  $ \langle E \rangle$ \\
\cline{1-9}
32 & 32 & 32 & $-$15.24881 &  &36 & 36 & 60 & $-$15.27363 \\
-- & -- & 36 & $-$15.26277 &  &40 & 40 &  -- & $-$15.27496 \\
-- & -- & 40 & $-$15.26604 &  &44 & 44 &  -- & $-$15.27521 \\
-- & -- & 44 & $-$15.26684 &  &46 & 46 &  -- & $-$15.27525 \\
-- & -- & 48 & $-$15.26703 &  &48 & 48 &  -- & $-$15.27526 \\
-- & -- & 52 & $-$15.26707 &  &50 & 50 & -- & $-$15.27527 \\
-- & -- & 56 & $-$15.26708 &  &52 & 52 & -- & $-$15.27527 \\
-- & -- & 60 & $-$15.26708 &  &54 & 54 & -- & $-$15.27527 \\
-- & -- & 64 & $-$15.26708 &  &56 & 56 & -- & $-$15.27527 \\
\cline{1-9}
\end{tabular}
\begin{tabbing}
$^{\dagger}$SNR value: $-$15.27527 a.u.  \hspace{50pt} \= 
$^{\ddagger}$Energy from GAMESS package \cite{schmidt93}: $-$15.27528 a.u. \\
\end{tabbing}
\end{table}
\endgroup

The present calculations employ following effective core potential (ECP) basis sets: SBKJC \citep{stevens84} for species 
containing Group-II elements and LANL2DZ \citep{hay85c} for Group-III or higher group elements. These are adopted from EMSL 
Basis Set Library\citep{feller96}. All one-electron integrals are generated by standard recursion relations \citep{obara86}
using Cartesian Gaussian-type orbitals as primitive basis functions. The norm-conserving pseudopotential matrix elements in 
contracted basis are imported from GAMESS\citep{schmidt93} suite of program package. The relevant LDA- and GGA-type of 
functionals in connection with B3LYP, BHLYP and PBE0 are: (i) Vosko-Wilk-Nusair (VWN)--with the homogeneous electron gas 
correlation proposed in parametrization formula V \citep{vosko80} (ii) B88--incorporating Becke\citep{becke88a} semi-local 
exchange (iii) Lee-Yang-Parr (LYP) \citep{lee88} semi-local correlation (iv) Perdew-Burke-Ernzerhof (PBE)\citep{perdew96} 
functional for semi-local exchange and correlation. All these are adopted from density functional repository 
program\citep{repository} except LDA. The convergence criteria imposed in this communication are slightly 
tighter than our earlier work\citep{roy08, roy08a, roy10, roy11}; this is to generate a more accurate orbital 
energies. Changes in following quantities were followed during the self-consistent field process, \emph{viz.,} (i) 
orbital energy difference between two successive iterations and (ii) absolute deviation in a density matrix element. They both 
were required to remain below a prescribed threshold set to $10^{-8}$ a.u.; this ensured that total energy maintained a 
convergence of at least this much, in general. In order to perform discrete Fourier transform, standard FFTW3 
package\citep{fftw05} was invoked. The resulting generalized matrix-eigenvalue problem is solved through standard LAPACK 
routine\citep{anderson99} accurately and efficiently. All molecular calculations are performed in their experimental 
geometries, taken from NIST database\citep{johnson16}, excepting those in Table~VI (see discussion later). Other 
details including scaling properties may be found in references \citep{roy08, roy08a, roy10, roy11, ghosal16, ghosal18}.

\begingroup                      
\begin{table}[t]      
\caption{\label{tab:table2} Convergence of KS energy$^{{\dagger},\ddagger}$ of Cl$_2$ ($R=4.2$ a.u.) in the grid ($h_r=0.3$) 
using B3LYP XC functional. All results are in a.u.}
\begin{tabular} {ccclcccrl}
\cline{1-9}
 \multicolumn{4}{c}{Set~I }  & & \multicolumn{4}{c}{Set~II}      \\
\cline{1-4} \cline{6-9} 
$N_x$   &   $N_y$   &    $N_z$  &  $ \langle E \rangle $ &    & $N_x$   &   $N_y$   &    $N_z$  &  $ \langle E \rangle$ \\
\cline{1-9}
32 & 32 & 32 & $-$28.49544 &  &36 & 36 & 76 & $-$29.78914 \\
-- & -- & 40 & $-$29.59219 &  &40 & 40 &  -- & $-$29.79282 \\
-- & -- & 48 & $-$29.75311 &  &44 & 44 &  --  & $-$29.79363 \\
-- & -- & 52 & $-$29.76710 &  &48 & 48 &  --  & $-$29.79380 \\
-- & -- & 56 & $-$29.77140 &  &52 & 52 &  -- & $-$29.79384 \\
-- & -- & 68 & $-$29.77304 &  &56 & 56 &  -- & $-$29.79384 \\
-- & -- & 72 & $-$29.77306 &  &60 & 60 &  -- & $-$29.79384 \\
-- & -- & 76 & $-$29.77306 & &64 & 64 & -- & $-$29.79384 \\
-- & -- & 80 & $-$29.77306 &  &68 & 68 &  -- & $-$29.79384 \\
\cline{1-9}
\end{tabular}
\begin{tabbing}
$^{\dagger}$SNR value: $-$29.79384 a.u.  \hspace{50pt} \= 
$^{\ddagger}$Energy from GAMESS package \cite{schmidt93}: $-$29.79390 a.u. \\
\end{tabbing}
\end{table}
\endgroup

\section{4. Result and Discussion}
Before proceeding for main results, at first it may be appropriate to discuss the grid optimization (in CCG) in 
terms of the RS parameter, $\zeta_{\mathrm{opt}}$. This is illustrated through convergence of total energy, as delineated in 
Eq.~(23). Table~1 shows this for a diatomic molecule HCl at its experimental geometry 
(bond length $1.275$\textup{\AA}). The HF energies are provided in non-uniform grid 
with respect to \emph{sparsity} (regulated by $N_x, N_{y}, N_{z}$) for a fixed grid spacing (determined by $h_r$, chosen as 
0.3), employing the NR and SNR schemes. The presentation strategy is similar to that in our 
previous works \citep{ghosal16, ghosal18}; accordingly we first vary $N_z$, the number of grid points along inter-nuclear axis, 
keeping the same along $xy$ plane static at certain reasonable value, say $N_x=N_y=32$. As $N_z$ is gradually increased from 32 
to 64 with an increment of 4, there is a smooth convergence in energy at around $N_z=56$ with a difference in total energy 
(we term it as grid accuracy) of about $ 10^{-6}$ a.u., between two successive steps. In the beginning, when $N_z$ goes 
through 40-44-48-52, one notices slow improvement in energy; after that it eventually attains the convergence for $N_z$ at around 
60. Then in Set~II in right-hand side, we vary $N_{x}, N_{y}$ along $xy$ plane keeping $N_z$ fixed at its previously determined 
value of $60$. Now we can see that convergence in energy takes place at $N_{x}=N_{y}=50$ with same grid accuracy of Set~I. 
Next, we have repeated the same calculation through the SNR scheme (ESP integral was approached analytically using recursion 
relations \citep{obara86} containing expensive incomplete Gamma functions); the corresponding electronic energy value is supplied 
in footnote of this table. For sake of completeness, energy is also quoted from GAMESS \cite{schmidt93} in footnote. It 
is quite encouraging to note that the two routes provide completely 
identical results as that in the reference. A similar exercise was undertaken for Cl$_2$ ($R_{eq}=4.2$ a.u.), and the 
conclusion remains same, i.e., converged total energies in this case were found to be as follows: $-$29.34810 (NR), $-$29.34809 
(SNR), $-$29.34813 (GAMESS) a.u. Thus it is evident that our direct, NR approach using RS-FCT in CCG, can correctly estimate the 
optimized RS-parameter, $\zeta_{\mathrm{opt}}$, and consequently, HF energy.

\begin{table}[t]     
\caption{\label{tab:table3} HF, B3LYP PBE0 energies (a.u.) of selected atoms. 
$E_{\mathrm{diff}}=|E{\mathrm{NR}}-E_{\mathrm{SNR}}|$.}
\resizebox{\textwidth}{90pt}{
\begin{tabular} {l|ccc|ccc|ccc}
\cline{1-10}
Atom & \multicolumn{9}{c}{$-\langle E \rangle$} \\
\cline{2-10}
  & \multicolumn{3}{c}{HF} & \multicolumn{3}{c}{B3LYP} & \multicolumn{3}{c}{PBE0} \\
 \cline{2-4}  \cline{5-7}  \cline{8-10}
  & NR & SNR & E$_{\mathrm{diff}}$ & NR & SNR& E$_{\mathrm{diff}}$ 
& NR & SNR  & E$_{\mathrm{diff}}$ \\ 
\cline{1-10}
Be & 0.96019 & 0.96019 & 0.00000 & 0.99386 & 0.99386 & 0.00000 & 0.99598 & 0.99598 & 0.00000 \\
B  & 2.53426 & 2.53426 & 0.00000 & 2.59552 & 2.59552 & 0.00000 & 2.59751 & 2.59751 & 0.00000 \\
C  & 5.30354 & 5.30354 & 0.00000 & 5.39301 & 5.39301 & 0.00000 & 5.39603 & 5.39603 & 0.00000 \\
N  & 9.61972 & 9.61969 & 0.00003 & 9.73283 & 9.73282 & 0.00001 & 9.74190 & 9.74189 & 0.00001 \\
O  &15.61720 &15.61681 & 0.00039 &15.80556 &15.80549 & 0.00007 &15.80351 &15.80341 & 0.00010 \\
Al & 1.86947 & 1.86947 & 0.00000 & 1.92379 & 1.92379 & 0.00000 & 1.93075 & 1.93075 & 0.00000 \\
Si & 3.67570 & 3.67570 & 0.00000 & 3.75077 & 3.75077 & 0.00000 & 3.76281 & 3.76281 & 0.00000 \\
P  & 6.31523 & 6.31523 & 0.00000 & 6.40580 & 6.40580 & 0.00000 & 6.42550 & 6.42550 & 0.00000 \\
S  & 9.87464 & 9.87464 & 0.00000 &10.01827 &10.01827 & 0.00000 &10.03519 &10.03519 & 0.00000 \\
Cl &14.68130 &14.68129 & 0.00001 &14.87170 &14.87170 & 0.00000 &14.88873 &14.88873 & 0.00000 \\
Ga & 1.94526 & 1.94526 & 0.00000 & 2.00195 & 2.00195 & 0.00000 & 2.00831 & 2.00831 & 0.00000 \\
Ge & 3.59814 & 3.59814 & 0.00000 & 3.67466 & 3.67466 & 0.00000 & 3.68693 & 3.68693 & 0.00000 \\
As & 5.95615 & 5.95615 & 0.00000 & 6.04808 & 6.04808 & 0.00000 & 6.06823 & 6.06823 & 0.00000 \\
Se & 9.01108 & 9.01108 & 0.00000 & 9.15439 & 9.15439 & 0.00000 & 9.17235 & 9.17235 & 0.00000 \\
Br &12.91872 &12.91872 & 0.00000 &13.10652 &13.10652 & 0.00000 &13.12535 &13.12535 & 0.00000 \\\cline{1-10}
\end{tabular} }
\end{table}

Next we move towards implementation of hybrid functional, B3LYP using above procedure. For this, we again consider a 
closed shell molecule, such as chlorine (Cl$_2$) at its experimental bond length of $4.2$ a.u., along $z$ axis as a 
specimen case. The formal convergence and stability of solution is illustrated in Table~2 at a grid spacing of 
$h_{r} = 0.3$. Same convergence criteria of previous table were imposed here as well. Once again the NR, SNR (provided 
in footnote) and GAMESS (also in footnote) energies excellently match with each other. 
An analogous study on HCl produced same result in energy, i.e., converged values in this case were $-$15.50844 (NR), 
$-$15.50843 (SNR), $-$15.50845 (GAMESS) a.u. This therefore establishes that our simple RS-FCT scheme 
in CCG can produce correct, meaningful results for XC functionals containing HF exchange. 

In order to extend the scope and applicability of proposed scheme, we now report total energies for a set of 15 atoms and 
15 molecules (along with test molecular systems) in Tables~3 and 4 respectively. To put things in perspective, we compare 
the NR and SNR results for three sets of calculations, namely HF, B3LYP and PBE0. Both tables engage identical basis set, 
pseudopotential and convergence (both grid and SCF) criteria. In all occasions, the maximum absolute deviation in energy (labeled 
$\mathrm{E}_{\mathrm{diff}}$) between NR and SNR, are produced side by side for easy comparison. The overall agreement between 
these two sets of results are in general excellent. A closer observation at Table~3 reveals that, the two energies are virtually 
indistinguishable for all the species with exception of O atom, where absolute deviation remains well below 
0.0004 a.u. Similarly in Table~4 again, these two results coincide for 13 molecules; in this case the largest discrepancy
(0.0001 a.u.) occurs for CH$_4$. Note that, since the previous tables have established that NR and SNR energies are very close to 
reference GAMESS results (which have also been verified in these cases), we do not reproduce those energies any more.  
   
\begin{table}[t]     
\caption{\label{tab:table4} HF, B3LYP PBE0 energies (a.u.) of molecules. 
$E_{\mathrm{diff}}=|E{\mathrm{NR}}-E_{\mathrm{SNR}}|$.}
\resizebox{\textwidth}{70pt}{
\begin{tabular} {lccc|ccc|ccc}
\cline{1-10}
Molecule & \multicolumn{9}{c}{$-\langle E \rangle$} \\
\cline{2-10}
  & \multicolumn{3}{c}{HF} & \multicolumn{3}{c}{B3LYP} & \multicolumn{3}{c}{PBE0} \\
 \cline{2-4}  \cline{5-7}  \cline{8-10}
  & NR & SR & E$_{\mathrm{diff}}$ & NR & SR & E$_{\mathrm{diff}}$ & NR & SR & E$_{\mathrm{diff}}$ \\ 
\cline{1-10}
Cl$_2$ & 29.34810 & 29.34809 & 0.00001 & 29.79384 & 29.79384 & 0.00001 & 29.82896 & 29.82896 & 0.00000 \\
Br$_2$ & 25.82400 &  25.82400 & 0.00000 & 26.25227 & 26.25227 & 0.00000 & 26.29481 & 26.29481 & 0.00000 \\
I$_2$  & 22.30720 & 22.30720 & 0.00000 & 22.71498 & 22.71498 & 0.00000 & 22.76592 & 22.76592 & 0.00000 \\
HCl    & 15.27527 & 15.27527 & 0.00000 & 15.50844 & 15.50843 & 0.00001 & 15.52803 & 15.52803 & 0.00000 \\
HBr    & 13.49506 & 13.49506 & 0.00000 & 13.72537 & 13.72537 & 0.00000 & 13.74603 & 13.74603 & 0.00000 \\
HI     & 11.72279 & 11.72279 & 0.00000 & 11.94619 & 11.94619 & 0.00000 & 11.96939 & 11.96939 & 0.00000 \\
H$_2$S & 11.02605 & 11.02605 & 0.00000 & 11.25511 & 11.25511 & 0.00000 & 11.27394 & 11.27394 & 0.00000 \\
H$_2$Se  & 10.14548 & 10.14548 & 0.00000 & 10.37371 & 10.37371 & 0.00000 & 10.39215 & 10.39215 & 0.00000 \\
PH$_3$   & 8.01636 & 8.01636 & 0.00000 & 8.23350 & 8.23350 & 0.00000 & 8.24942 & 8.24942 & 0.00000 \\
CH$_3$Cl & 21.87306 & 21.87306 & 0.00000 & 22.29768 & 22.29768 & 0.00000 & 22.33525 & 22.33525 & 0.00000\\
CH$_4$   & 7.78878 & 7.78888 & 0.00010 & 8.00843 & 8.00846 & 0.00003 & 8.02684 & 8.02686 & 0.00002 \\
SiH$_3$Cl & 20.19863 & 20.19862 & 0.00001 & 20.58442 & 20.58441 & 0.00001 & 20.61885 & 20.61885 & 0.00000\\
Si$_2$H$_6$ & 10.93377 & 10.93377 & 0.00000 & 11.28249 & 11.28249 & 0.00000 & 11.31207 & 11.31207 & 0.00000 \\
P$_4$ & 25.13843 & 25.13843 & 0.00000 & 25.81974 & 25.81974 & 0.00000 & 25.91452 & 25.91452 & 0.00000 \\
SiH$_4$ & 6.03289 & 6.03289 & 0.00000 & 6.22621 & 6.22621 & 0.00000 & 6.23634 & 6.23634 & 0.00000 \\
\cline{1-10}
\end{tabular}}
\end{table}

In the last part of this study, in Table~5, we investigate the highest occupied molecular orbital (HOMO) energies of same 
set of atoms and molecules of Tables 3 and 4. As expected from previous tables, NR and SNR  
schemes produce practically identical results in these cases too. Hence now onwards, we present 
only NR results to save space. The available theoretical and experimental IPs are recorded for easy comparison.  
It is revealed that even though the theory does not account for electron correlation and relaxation effects, HF
HOMO energies still compare very favorably with experiment than any of the three DFT functionals considered. This is apparent 
from the respective deviations; one notices underestimation of 14-41\%, 9-37\%, 1-28\% for B3LYP, PBE0 and BHLYP functionals, 
whereas HF results show absolute deviation by 0-14\%. However, it is interesting to note that, amongst the three functionals, the 
deviations fall from B3LYP to BHLYP, mainly because the fractional content of HF exchange (which has a pivotal role in 
determining correct asymptotic nature at long range) gradually increases in that order. This observation along with a 
consideration of total energies (from Tables~3 and 4), suggests that hybrid functionals provide a systematic improvement 
over the HF result. These observations are 
further complemented in Table~6, where a similar analysis is made for some $\pi$-electron molecules (simple, conjugated as well
as aromatic). A very important aspect of such species is the so-called fundamental gap, which is the difference in energy 
between HOMO and LUMO. A satisfactory estimation of this gaps requires accurate knowledge of these two orbital energies.  
The current table is an attempt towards this direction. In this case, we have taken the calculated geometry (using B3LYP XC 
functional and cc-PVTZ basis set) from NIST \cite{johnson16} data base. As expected, we witness similar kind of pattern in 
calculated IP values as in previous table; with respect to experimental/theoretical values, the range of underestimation for hybrid 
functionals are: 18-32\%, 15-28\% and 11-19\% respectively, for B3LYP, PBE0 and BHLYP. 
 
\begingroup
\begin{table}[tp]     
\caption{\label{tab:table5} Negative HOMO energies, $-\epsilon_{\mathrm{HOMO}}$ (in a.u.) for selected atoms and molecules 
using HF, B3LYP, PBE0 and BHLYP XC functionals.}
\resizebox{\textwidth}{70pt}{
\begin{tabular} {lccccccccccc}
\cline{1-12}
Atom & \multicolumn{5}{c}{$-\epsilon_{\mathrm{HOMO}}$(a.u.)} & Molecule & \multicolumn{5}{c}{$-\epsilon_{\mathrm{HOMO}}$(a.u.)} \\
\cline{2-6} \cline{8-12}
  & HF & B3LYP & PBE0 & BHLYP & Expt.$^{\dagger}$ & & HF & B3LYP & PBE0 & BHLYP & Expt.$^{\ddagger}$ \\ 
\cline{1-12}
Be & 0.3090 & 0.2291 & 0.2387 & 0.2650 & 0.3426 & Cl$_2$ & 0.4786 & 0.3274 & 0.3433 & 0.3947 & 0.4219 \\
B  & 0.3080 & 0.1830 & 0.1929 & 0.2348 & 0.3050 & Br$_2$ & 0.4139 & 0.2889 & 0.3034 & 0.3456 & 0.3864 \\
C  & 0.4313 & 0.2615 & 0.2771 & 0.3290 & 0.4138 & I$_2$ & 0.3695 & 0.2658 & 0.2793 & 0.3138 & 0.3429 \\
N  & 0.5657 & 0.3497 & 0.3707 & 0.4338 & 0.5341 & HCl & 0.4772 & 0.3274 & 0.3425 & 0.3937 & 0.4700 \\
O  & 0.5098 & 0.3247 & 0.3379 & 0.4090 & 0.5005 & HBr & 0.4302 & 0.3030 & 0.3170 & 0.3602 & 0.4303 \\
Al & 0.2095 & 0.1291 & 0.1408 & 0.1621 & 0.2200 & HI & 0.3875 & 0.2804 & 0.2932 & 0.3295 & 0.3815 \\
Si & 0.3029 & 0.1937 & 0.2088 & 0.2374 & 0.3000 & H$_2$S & 0.3895 & 0.2617 & 0.2746 & 0.3191 & 0.3851 \\
P  & 0.3901 & 0.2545 & 0.2732 & 0.3076 & 0.3854 & H$_2$Se & 0.3613 & 0.2466 & 0.2593 & 0.2987 & 0.3649 \\
S  & 0.3631 & 0.2506 & 0.2625 & 0.3043 & 0.3807 & PH$_3$ & 0.3849 & 0.2675 & 0.2781 & 0.3200 & 0.3626 \\
Cl & 0.4731 & 0.3294 & 0.3442 & 0.3944 & 0.4766 & CH$_3$Cl & 0.4340 & 0.2946 & 0.3084 & 0.3567 & 0.4149 \\
Ga & 0.2058 & 0.1263 & 0.1381 & 0.1590 & 0.2205 & CH$_4$ & 0.5416 & 0.3882 & 0.4013 & 0.4555 & 0.4998 \\
Ge & 0.2844 & 0.1825 & 0.1974 & 0.2232 & 0.2903 & SiH$_3$Cl & 0.4509 & 0.3149 & 0.3288 & 0.3754 & 0.4281 \\
As & 0.3665 & 0.2426 & 0.2605 & 0.2910 & 0.3607 & Si$_2$H$_6$ & 0.4068 & 0.3043 & 0.3152 & 0.3516 & 0.3870 \\
Se & 0.3319 & 0.2337 & 0.2451 & 0.2815 & 0.3584 & P$_4$ & 0.3844 & 0.2921 & 0.3075 & 0.3362 & 0.3381 \\
Br & 0.4206 & 0.3010 & 0.3144 & 0.3563 & 0.4341 & SiH$_4$ & 0.4871 & 0.3576 & 0.3687 & 0.4149 & 0.4520 \\
\cline{1-12}
\end{tabular}}
\begin{tabbing}
$^{\dagger}$Optical spectroscopy\citep{johnson16}. \= \hspace{170pt}
$^{\ddagger}$Photo-electron spectroscopy \citep{johnson16}. 
\end{tabbing}
\end{table}
\endgroup

A few remarks may now be made before passing. During the calculation, it is realized that the computational burden of our NR route is 
quite less (about eight times) than the SNR one. It occurs because of the obvious reason that the former route obviates the 
necessity of two primitive loops in Eq.~(3); instead the ESP integral is directly calculated in CCG using contracted Gaussian 
functions. In order to make the SNR route attractive and fruitful for practical applications, an elegant solution was proposed 
\citep{liu17} in the form of some optimal 
recurrence relations for analytical evaluation of ESP integral \citep{liu17}. We have not attempted to incorporate this 
procedure in this communication, as this is aside the main theme and purpose of this work. Here, we just wanted to establish the 
validity, efficacy and feasibility of a simple direct NR scheme in the broader context of certain DFT functionals for 
large-scale calculations. As demonstrated in this work, the NR route is quite efficient, and may be favorable, especially
for larger basis set. The accuracy and ease of implementation augurs well for its further application in the development 
of range-separated hybrid functional from \emph{first principles}. This leads to a more complete representations of HF exchange 
in asymptotic region which closes the gap between theoretical and experimental result, keeping the computational overhead at same 
level as in commonly used hybrid functionals\citep{kronik12}, and currently we are working on it.

\begin{table}[tp]     
\caption{\label{tab:table6}  Negative HOMO energies, $-\epsilon_{\mathrm{HOMO}}$ (in a.u.) for selected $\pi$-electron molecules 
using HF, B3LYP, PBE0 and BHLYP XC functionals.}
\begin{tabular} {lcccccc}
\cline{1-7}
Molecule & \multicolumn{6}{c}{$-\epsilon_{\mathrm{HOMO}}$(a.u.)} \\
\cline{2-7} 
	& HF & B3LYP & PBE0 & BHLYP & Theory\citep{johnson16} & Expt.$^{\dagger}$ \\ 
\cline{1-7}
Ethylene & 0.3686 & 0.2649 & 0.2796 & 0.3140 & 0.376$^a$ & 0.3859 \\
Propene & 0.3544 & 0.2503 & 0.2645 & 0.2994 & 0.354$^a$ & 0.3565 \\
1,3-Butadiene (E) & 0.3188 & 0.2308 & 0.2444 & 0.2734 & 0.332$^a$ & 0.3333 \\
1,3-Pentadiene (E) & 0.3053 & 0.2175 & 0.2309 & 0.2600 & 0.306$^b$ & 0.3157 \\
1,3,5-Hexatriene (E) & 0.2862 & 0.2090 & 0.2218 & 0.2471 & 0.294$^c$ & 0.3050 \\
1,3-Cyclo-pentadiene & 0.3051 & 0.2142 & 0.2283 & 0.2584 & 0.301$^d$ & 0.3135 \\
Benzene & 0.3362 & 0.2491 & 0.2637 & 0.2920 & 0.329$^e$ & 0.3399 \\
Thiophene & 0.3338 & 0.2400 & 0.2544 & 0.2856 & 0.327$^f$ & 0.3271 \\
Acetylene & 0.4110 & 0.2928 & 0.3073 & 0.3473 & 0.409$^a$ & 0.4189 \\
Acetaldehyde & 0.4628 & 0.3072 & 0.3193 & 0.3764 & 0.425$^a$ & 0.3758 \\
\cline{1-7}
\end{tabular}
\begin{tabbing}
\hspace*{9cm}\= \kill
$^{\dagger}$Photo-electron spectroscopy \citep{johnson16}. \> 
$^a$CCSD result using cc-PVTZ basis. \\ 
$^b$CCD result using 6-31G$^\star$ basis.  \>
$^c$QCISD result using 6-311G$^{\star \star}$ basis. \\
$^d$CCD result using 6-31G$^{\star \star}$ basis. \>
$^e$CCSD result using 6-31G$^{\star \star}$ basis. \\
$^e$CCSD result using 6-311G$^{\star \star}$ basis.
\end{tabbing}
\end{table}

\section{4. Future and Outlook}
We have demonstrated the feasibility and practicability of a direct NR scheme for computation of HF exchange energy 
density, energy and matrix in real-space CCG. This was applied for a host of atoms and molecules; properties such as total 
energy and orbital energies were offered, within a pseudopotential KS-DFT. Besides HF components, these were also presented 
for B3LYP, PBE0 and BHLYP hybrid functionals. The success of 
this approach relies on accurate estimation of ESP integral, which in turn depends on optimization of the RS 
parameter in Coulomb kernel. The obtained results for all 
these species from both NR and SNR scheme, are in excellent agreement with each other. A consideration of scaling 
properties suggests that this approach may be quite helpful in large-scale DFT calculations involving orbital-dependent
functionals. Application to range-separated hybrid, hyper as well as local hybrid XC functionals would further enhance 
its scope in a wide range of systems. It may also be desirable to examine its performance in various important physio-chemical 
properties of many-electron systems. 

\section{Acknowledgement}
AG is grateful to UGC for a Senior Research Fellowship (SRF). TM acknowledges IISER Kolkata for a Junior Research 
fellowship (JRF). Financial support from DST SERB, New Delhi, India (sanction order number EMR/2014/000838) is gratefully 
acknowledged.
\bibliography{dftbib.bib}
\end{document}